# Hollow core Bragg fiber with antiresonant intermediate layer

Yurii A. Zinin, Andrey V. Panov, Yurii N. Kulchin

Institute of Automation and Control Processes, Far Eastern Branch of Russian Academy of Sciences, 5, Radio st., Vladivostok, 690041, Russia

## ABSTRACT

By means of the transfer matrix method, the optical properties of fibers with a distinct intermediate layer between a hollow core and periodic cladding are calculated. The periodic cladding consists of two types of the alternating layers. The intermediate layer has distinct thickness and refractive index. Depending on these parameters, the fiber can work in the single-mode or multi-mode regimes. In the multi-mode regime, the optical loss of the smallest loss mode can be decreased by increasing the thickness of the layer. In the single-mode regime, the optical loss falls with a rise in the refractive index of the intermediate layer. The optical properties of the fiber are determined by the antiresonance reflection from the intermediate layer and the Bragg reflection from the periodic cladding. Selecting the parameters of the intermediate layer, the optical loss of the fiber in the single-mode regime can be reduced by an order of magnitude over the loss of the traditional Bragg fiber.

## 1. INTRODUCTION

In the Bragg fibers the guiding of light is determined by the Bragg reflection from periodic cladding layers [1]. The idea to use leakage modes in periodic dielectric waveguides was first proposed in Ref. [2]. The leakage loss differs substantially for various modes with no polarization degeneracy [3, 4]. Because of the guiding principle, the Bragg fiber can be considered as a kind of the photonic bandgap fiber [5]. So, there has been a large interest in studying the Bragg fibers.

When the period of binary layered claddings exceeds the light wavelength, the mechanism of transmission in the waveguide changes from the Bragg to the antiresonant. The optical properties of such fibers are determined mainly by the first, nearest to the core, cladding layer [6–8]. Rowland et al. [9] put forward a model accounting for antiresonances in the Bragg fibers. Equivalence between the in-phase (generalized quarter-wave stack) and antiresonant reflection conditions in the Bragg fiber was shown in Ref. [10].

Duguay et al. [11] first studied the antiresonances in the planar waveguide comprising several layers with different parameters. Later, Archambault et. al. [12] in addition to planar geometry also examined cylindrical waveguides with diverse layers. Although, the original definition of the Bragg fiber implies the purely periodic cladding structure, the properties of the Bragg reflection fibers with the distinct intermediate layer has already been examined before. Mizrahi and Schächter [13] called this layer as "matching" whose specific thickness allows the Bragg fiber to support a symmetric mode with a specified core field distribution. In Ref. [14] the optical properties of the similar fiber were investigated with special emphasis on the influence of the thickness of the intermediate layer between the core and periodic cladding. In this study, we investigate the effect of both the thickness and the refractive index of this layer on the optical properties of the Bragg fiber with the distinct intermediate layer under the assumption of non-absorbing fiber material.

## 2. GEOMETRY OF BRAGG FIBER WITH INTERMEDIATE LAYER

Fig. 1 represents the profile of the Bragg fiber with the distinct intermediate layer. Unlike the traditional Bragg fiber with the completely periodic cladding with the alternating two types of layers, the layer adjacent to the core has the specific thickness $d_M$ and refractive index $n_M$. We calculate the optical properties of such fibers by means of the transfer matrix method [14].

For an any homogeneous cylindrical layer, the axial and azimuthal components of the electromagnetic field ($E_z$, $E_\phi$, $H_z$, $H_\phi$) can be expressed as a linear combination of two Hankel functions of the first and second kinds; $H_m^I(kr)$ and $H_m^{II}(kr)$. The electromagnetic field components in the *l*th layer can be written as (here the exp(*iβz*) factor, common to all the field components, is omitted):

$$E_z = \left[ A_l H_m^I(k_l r) + B_l H_m^{II}(k_l r) \right] \begin{Bmatrix} \cos m\phi \\ \sin m\phi \end{Bmatrix},$$

$$E_\phi = \frac{i\beta m}{r k_l^2} \left[ A_l H_m^I(k_l r) + B_l H_m^{II}(k_l r) \right] \begin{Bmatrix} \sin m\phi \\ -\cos m\phi \end{Bmatrix} + \frac{i\omega\mu}{c k_l} \left[ C_l \dot{H}_m^I(k_l r) + D_l \dot{H}_m^{II}(k_l r) \right] \begin{Bmatrix} \sin m\phi \\ \cos m\phi \end{Bmatrix},$$

$$H_z = \left[ C_l H_m^I(k_l r) + D_l H_m^{II}(k_l r) \right] \begin{Bmatrix} \sin m\phi \\ \cos m\phi \end{Bmatrix},$$

$$H_\phi = -\frac{i\omega\varepsilon}{c k_l} \left[ A_l \dot{H}_m^I(k_l r) + B_l \dot{H}_m^{II}(k_l r) \right] \begin{Bmatrix} \cos m\phi \\ \sin m\phi \end{Bmatrix} - \frac{i\beta m}{r k_l^2} \left[ C_l H_m^I(k_l r) + D_l H_m^{II}(k_l r) \right] \begin{Bmatrix} \cos m\phi \\ -\sin m\phi \end{Bmatrix},$$

where $n_l$ is the refractive index of the *l*th layer and $k_l$ is the transverse wave vector component in the *l*th layer, $k_l^2 = k_0^2 (n_l^2 - \beta^2 / k_0^2)$, $k_0$ is the wave vector in vacuum, $\beta$ is the propagation constant of the guided mode, $\beta/k_0 = n_{ef}$ is the effective refractive index of the mode; $A_l$, $B_l$, $C_l$, $D_l$ are the amplitude constants for the *l*th layer.

The longitudinal electromagnetic field components must be continuous across the boundaries of layers. The boundary conditions can be expressed in matrix:

$$\begin{pmatrix} A_{l+1} \\ B_{l+1} \\ C_{l+1} \\ D_{l+1} \end{pmatrix} = M(r_l) \begin{pmatrix} A_l \\ B_l \\ C_l \\ D_l \end{pmatrix},$$

where $M(r_l)$ is a 4×4 matrix consisting of elements $m_{gq}$, $g$ and $q$ are the line and column. The matrix elements take the form:

$$m_{11} = \frac{1}{4} i\pi y_l \left\{ H_m^I(x_l) \dot{H}_m^{II}(y_l) - \frac{\varepsilon_l y_l}{\varepsilon_{l+1} x_l} \dot{H}_m^I(x_l) H_m^{II}(y_l) \right\},$$

$$m_{12} = \frac{1}{4} i\pi y_l \left\{ H_m^{II}(x_l) \dot{H}_m^{II}(y_l) - \frac{\varepsilon_l y_l}{\varepsilon_{l+1} x_l} \dot{H}_m^{II}(x_l) H_m^{II}(y_l) \right\},$$

$$m_{13} = \frac{i y_l m \beta \lambda_0}{8 \varepsilon_{l+1}} \left\{ \frac{1}{y_l} - \frac{y_l}{x_l^2} \right\} H_m^I(x_l) H_m^{II}(y_l),$$

$$m_{14} = \frac{i y_l m \beta \lambda_0}{8 \varepsilon_{l+1}} \left\{ \frac{1}{y_l} - \frac{y_l}{x_l^2} \right\} H_m^{II}(x_l) H_m^{II}(y_l),$$

$$m_{21} = \frac{1}{4} i\pi y_l \left\{ \frac{\varepsilon_l y_l}{\varepsilon_{l+1} x_l} \dot{H}_m^I(x_l) H_m^I(y_l) - H_m^I(x_l) \dot{H}_m^I(y_l) \right\},$$

$$m_{22} = \frac{1}{4} i\pi y_l \left\{ \frac{\varepsilon_l y_l}{\varepsilon_{l+1} x_l} \dot{H}_m^{II}(x_l) H_m^I(y_l) - H_m^{II}(x_l) \dot{H}_m^I(y_l) \right\},$$

$$m_{23} = \frac{i y_l m \beta \lambda_0}{8 \varepsilon_{l+1}} \left\{ \frac{y_l}{x_l^2} - \frac{1}{y_l} \right\} H_m^I(x_l) H_m^I(y_l),$$

$$m_{24} = \frac{i y_l m \beta \lambda_0}{8 \varepsilon_{l+1}} \left\{ \frac{y_l}{x_l^2} - \frac{1}{y_l} \right\} H_m^{II}(x_l) H_m^I(y_l),$$

$$m_{31} = \frac{1}{8} i y_l m \beta \lambda_0 \left\{ \frac{1}{y_l} - \frac{y_l}{x_l^2} \right\} H_m^I(x_l) H_m^{II}(y_l),$$

$$m_{32} = \frac{1}{8} i y_l m \beta \lambda_0 \left\{ \frac{1}{y_l} - \frac{y_l}{x_l^2} \right\} H_m^{II}(x_l) H_m^{II}(y_l),$$

$$m_{33} = \frac{1}{4} i \pi y_l \left\{ H_m^{I}(x_l) \dot{H}_m^{II}(y_l) - \frac{y_l}{x_l} \dot{H}_m^{I}(x_l) H_m^{II}(y_l) \right\},$$

$$m_{34} = \frac{1}{4} i \pi y_l \left\{ H_m^{II}(x_l) \dot{H}_m^{II}(y_l) - \frac{y_l}{x_l} \dot{H}_m^{II}(x_l) H_m^{II}(y_l) \right\},$$

$$m_{41} = \frac{1}{8} i y_l m \beta \lambda_0 \left\{ \frac{y_l}{x_l^2} - \frac{1}{y_l} \right\} H_m^{I}(x_l) H_m^{I}(y_l),$$

$$m_{42} = \frac{1}{8} i y_l m \beta \lambda_0 \left\{ \frac{y_l}{x_l^2} - \frac{1}{y_l} \right\} H_m^{II}(x_l) H_m^{I}(y_l),$$

$$m_{43} = \frac{1}{4} i \pi y_l \left\{ \frac{y_l}{x_l} \dot{H}_m^{I}(x_l) H_m^{I}(y_l) - H_m^{I}(x_l) \dot{H}_m^{I}(y_l) \right\},$$

$$m_{44} = \frac{1}{4} i \pi y_l \left\{ \frac{y_l}{x_l} \dot{H}_m^{II}(x_l) H_m^{I}(y_l) - H_m^{II}(x_l) \dot{H}_m^{I}(y_l) \right\},$$

where $x_l = k_l r_l$, $y_l = k_{l+1} r_l$, $\lambda_0$ is the vacuum wavelength.

The constants $A$, $B$, $C$, $D$ for the $(l + 1)$th layer of the cladding can be expressed through those for the core as:

$$\begin{pmatrix} A_{l+1} \\ B_{l+1} \\ C_{l+1} \\ D_{l+1} \end{pmatrix} = \prod_{i=1}^{l} M(r_i) \begin{pmatrix} A_1 \\ A_1 \\ C_1 \\ C_1 \end{pmatrix}. \tag{1}$$

Since, outside the waveguide, there must be no reflected wave, so the constants $B_{N+1}$ and $D_{N+1}$ for the fiber with $N$ layers are zero. A solution to Eq. (1) at $l = N$ exists under following condition

$$(p_{23} + p_{24})(p_{41} + p_{42}) - (p_{43} + p_{44})(p_{21} + p_{22}) = 0, \tag{2}$$

where $p_{gq}$ are the entries of the matrix obtained after multiplying all the $M(r_i)$ matrices. Eq. (2) makes it possible to find the modes of the multilayer cylindrical fiber. Given the azimuthal mode number is zero, Eq. (2) takes a simpler form. For $TM_{0k}$ modes, $k$ being the radial mode number, we obtain:

$$p_{21} + p_{22} = 0$$

For $TE_{0k}$ modes, the equation takes the form:

$$p_{43} + p_{44} = 0$$

The optical loss, $\gamma$, is defined as a reduction in the intensity of light propagating through the fiber: $\gamma = (4\pi \, \mathrm{Im} \, n_{ef})/\lambda_0$, or in dB m$^{-1}$ units,

$$\gamma = \frac{40\pi \lg e}{\lambda_0} \mathrm{Im} \, n_{ef}.$$

In this work, we study the fiber with the following parameters: the hollow core with radius $r_1$=1.8278 μm, periodic cladding consists of 15 layers with $d_H$=0.2133 μm, $n_H$=1.49 and 16 layers with $d_L$=0.346 μm, $n_L$=1.17; the parameters of the intermediate layer vary. Optical loss $\gamma$ is calculated for the radiation with the wavelength $\lambda$=1 μm. This structure of core and periodic cladding tends to the quarter-wave stack condition. In the specific case, when $d_M$=0.2133 μm and $n_M$=1.49, this fiber has been investigated repeatedly [4, 15, 16].

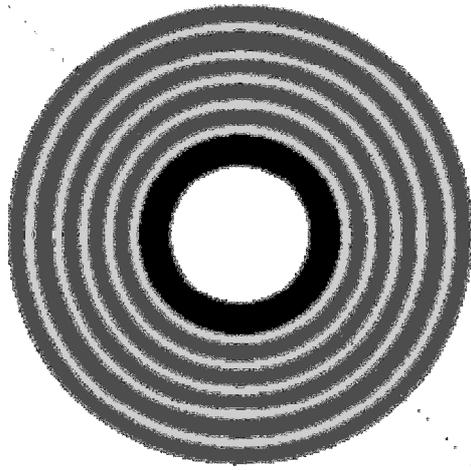

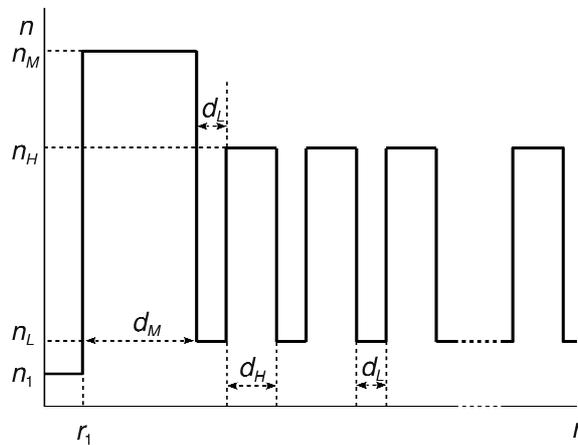

Figure 1. Schematic representation of the Bragg fiber with the intermediate layer.

## 3. EFFECT OF INTERMEDIATE LAYER PARAMETERS ON FIBER OPTICAL LOSS

Fig. 2 depicts the dependence of the fiber optical loss on the thickness and refractive index of the intermediate layer. The surface of the optical loss makes several wide valleys separated from each other by high ridges. For this fiber, within the examined range of $d_M$ and $n_M$, modes $TE_{01}$ and $TE_{02}$ have the lowest optical losses. After the computations with the transfer matrices, the optical loss at the valleys for all remaining modes proved to be several orders greater. In the neighborhood of ridges, $\gamma$ of some other modes exceed the loss of modes $TE_{01}$ and $TE_{02}$ only by an order. Here, the study is only limited to these two modes, since at the valleys remaining modes will not change the regime of guidance, and in the vicinity of ridges we will further speak about the multi-mode regime.

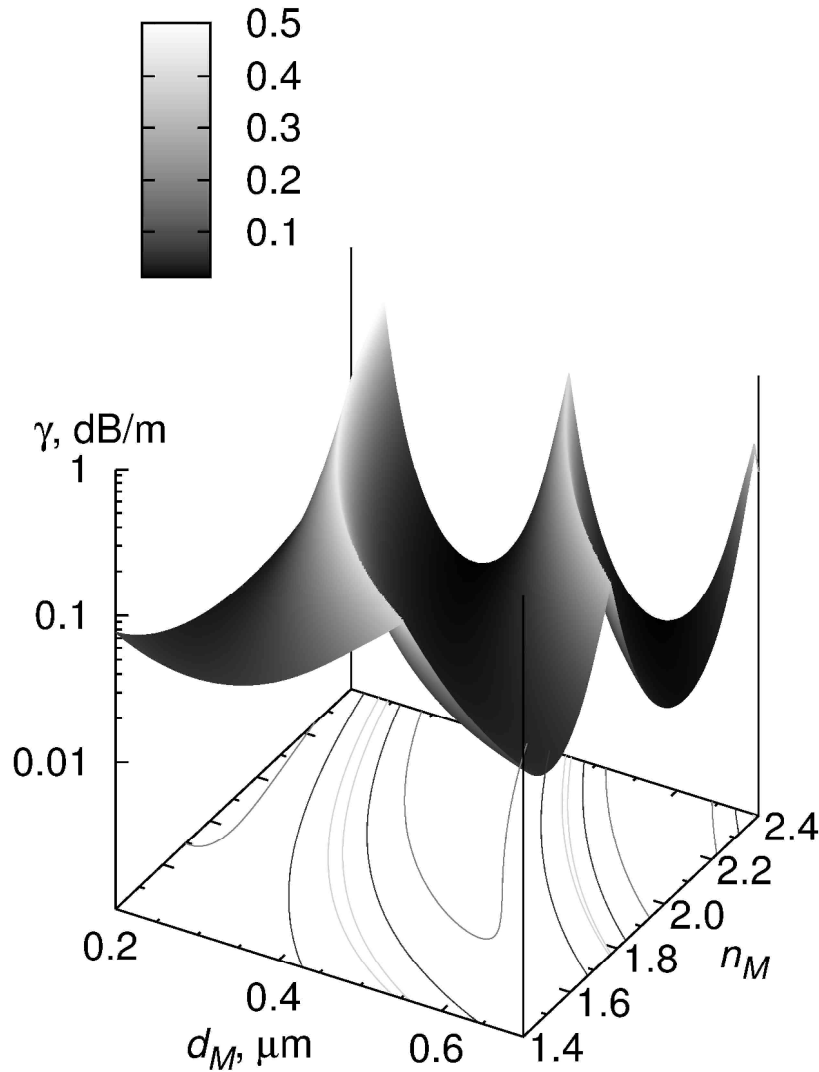

Figure 2. Dependence of the optical loss γ on the parameters of the intermediate layer.

In Fig. 2 we can clearly see two sharp ridges, the insignificant part of the third ridge is observed near the maximum values of $d_M$ and $n_M$ used for our calculations. On the base of the graph, the contour map with isolines of γ is depicted. The position of ridge crest corresponds precisely to $d_M$ and $n_M$ with the equal γ for modes $TE_{01}$ and $TE_{02}$. The larger the radius of the core of the traditional Bragg fiber, the higher the mode effective refractive index $n_{ef}$ increases, tending to the value of the core refractive index. The enhancement of $d_M$ and $n_M$ leads to the rise in effective refractive index, which corresponds to certain solution of the eigenvalue equation, however, since the core radius remains constant, the radial mode number begins to decrease. For the particular eigenvalue, the continuous curve (hereinafter referred to as the root curve), representing the effective refractive index or the optical loss as functions of $d_M$ or $n_M$, matches a set of modes with the fixed azimuthal and various radial numbers [14].

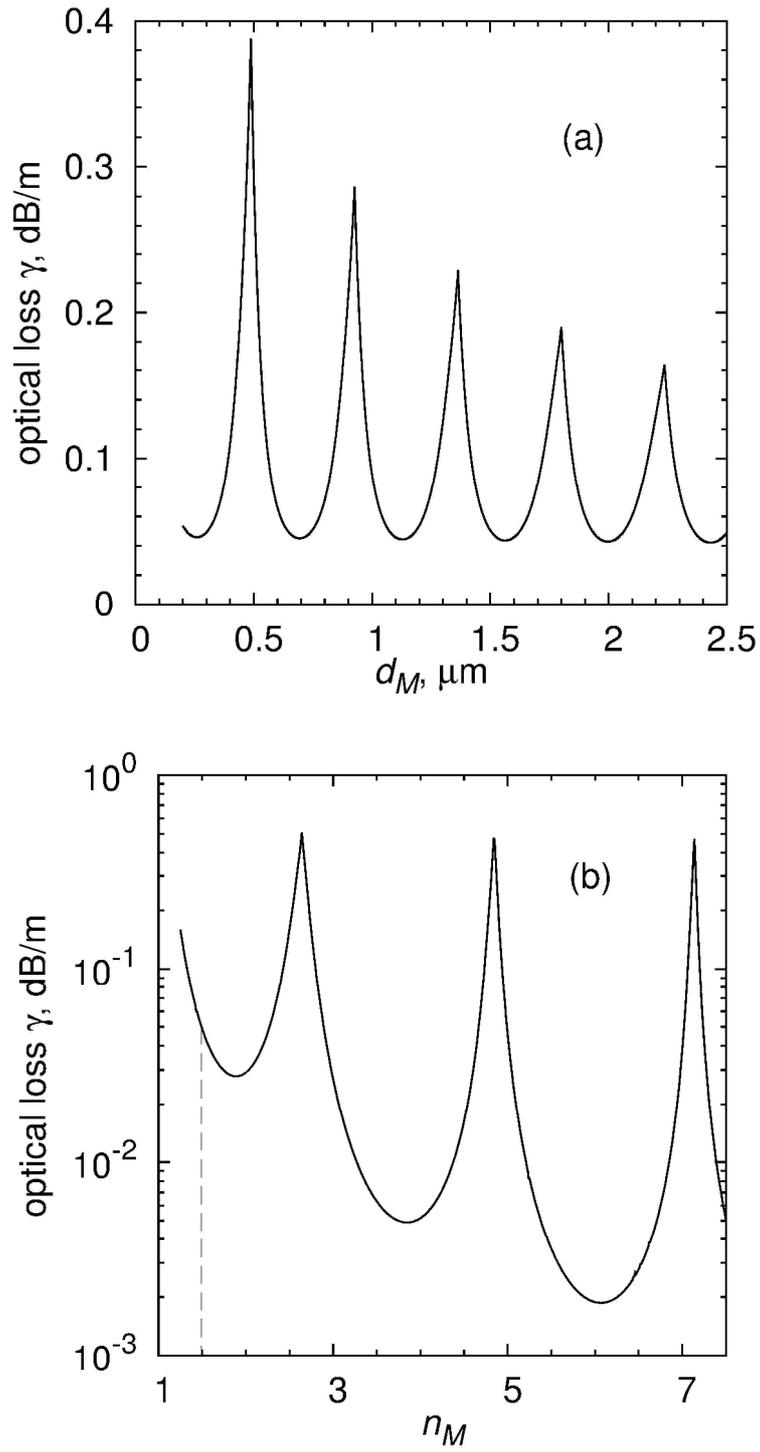

Figure 3. Dependence of the optical loss on thickness $d_M$ (a) and refractive index $n_M$ (b) of the intermediate layer with $n_M$=1.49 (a) or $d_M$=0.2133 μm (b). The minima (maxima) correspond to the single-mode (multi-mode) regime.

For each root curve associated with a set of $TE_{0k}$ modes, there exists roughly one and the same $n_{ef}$ with minimum $\gamma$ [14]. When $n_{ef}$ lies in the vicinity of this value, the waveguide has a transmission window. This window is due to the in-phase reflection condition [10]. By varying $d_M$ or $n_M$, each root curve falls within the transmission window only once. Thus, the change in the fiber transmission is caused by an incremental decrease in the radial number of the eigenvalue equation solution. If two eigenvalues occur simultaneously within the transmission window then, at some values of the intermediate layer parameters, the optical losses of these two root curves become equal; this corresponds to the ridge crests in Fig. 2. In our calculations, we always selected from all possible solutions of the eigenvalue equation the solution with the smallest $\gamma$. Because of very sharp ridges, this approximation introduces the insignificant error into the computation of the loss value only near the ridge crests. In this region, the fiber works in a multi-mode regime (at least two modes). These competing modes, which lead to the formation of the ridges, always are $TE_{01}$ and $TE_{02}$ (the left and the right sides of the ridges in Fig. 2, respectively).

Wide valleys of Fig. 2 correspond to the single-mode regime of light propagation (the $TE_{01}$ mode). In the descent along the right side of a ridge, the $TE_{02}$ mode transforms into the $TE_{01}$ mode. In intersecting a ridge crest along any trajectory, one solution of the eigenvalue equation transforms to another.

Fig. 3 shows $\gamma$ as functions of thickness $d_M$ (3.a) and refractive index $n_M$ (3.b) of the intermediate layer. As one can see from Fig. 3.a, with the growth of $d_M$, $\gamma$ has periodic oscillations, in this case the loss minimum magnitude remains virtually constant and the height of ridges decreases. In other words, in the multi-mode regime, $\gamma$ for the competing modes $TE_{01}$ and $TE_{02}$ decreases with an increase in $d_M$.

The dependence of $\gamma$ on $n_M$ exhibits two distinctive features: the optical loss oscillates with the rise in the refractive index and, in the single-mode regime, $\gamma$ reduces by virtue of an increase in the Fresnel reflection into the core. It is possible to lower substantially the loss by selection of the intermediate layer refractive index. As an example, for standard Bragg fiber B [15] $\gamma$=0.0503 dB/m (the dashed line in Fig. 3.b). If, at constant $d_M$=0.2133 μm, we increase the intermediate layer refractive index to the value of $n_M$=1.89 then the optical loss diminishes to 0.0279 dB/m. In Fig. 3.b, this corresponds to the first minimum of the optical loss. In this case, the fiber transmits the light with the wavelength of 1μm in the single-mode regime (the $TE_{01}$ mode). Further $n_M$ growth towards the maximum of $\gamma$ makes it possible the multi-mode light propagation. Then, the optical loss begins to fall again and in the second minimum of Fig. 3.b $\gamma$ reaches 0.0049 dB/m ($n_M$=3.85). As compared with standard fiber B, the optical loss drops by an order and the fiber works again in the single-mode regime. At $n_M$=6.07 there exists a third minimum of the optical loss: $\gamma$=0.0019 dB/m. The production of the intermediate dielectric layer with such a high refractive index appears to be difficult. The height of the ridges (maxima of $\gamma$) is virtually independent of $n_M$.

## 4. ANTIRESONANT PROPERTIES OF INTERMEDIATE LAYER

The solid lines in Fig. 4 are calculated with the transfer matrix method. These curves show valley bottom and ridge crest positions in the ($d_M$, $n_M$) plane where the maxima or minima of the optical loss are observed. The dotted lines near these curves match the condition of the existence of the resonance or antiresonance in the intermediate layer [8, 10, 12]

$$2k_{ex}d_M = \pi s, \qquad (3)$$

where $s$ is the number of the curve in Fig. 4, $k_{ex}$ is the transverse component of the propagation constant,

$$k_{ex} = \frac{2\pi}{\lambda}\sqrt{n_M^2 - n_{ef}^2} \ . \qquad (4)$$

The even values of $s$ match the resonance positions and odd $s$ correspond to the antiresonances. The propagation constant is computed using $n_{ef}$ obtained with transfer matrices. The curves of the resonances or antiresonances correspond to $s$ from 2 to 6. For $s$=1, the optical thickness of the intermediate layer is not sufficient to prevail antiresonant reflecting guidance mechanism in the transmission. Comparison of the curves in Fig. 4 shows that the resonance-antiresonance condition (3) corresponds better to the maxima of the optical loss than it does to the minima. As a possible explanation for this phenomenon we propose the following considerations. Near the maxima of the optical loss the variations of $\gamma$ due to changes in $d_M$ or $n_M$ are high, so that less significant additional contributions from other effects cannot substantially alter the position of the maxima. The wide valleys with insignificant variation in the magnitude of the loss match the minima of $\gamma$. Therefore, the additional minor disturbances can considerably move the minimum curve from the antiresonance condition. The additional perturbations may arise from different reasons: (i) the number of layers of the periodic cladding is finite; (ii) there is an aperiodicity of the cylindrical functions at their small arguments; (iii) other layers may cause the additional resonance structure of the fiber.

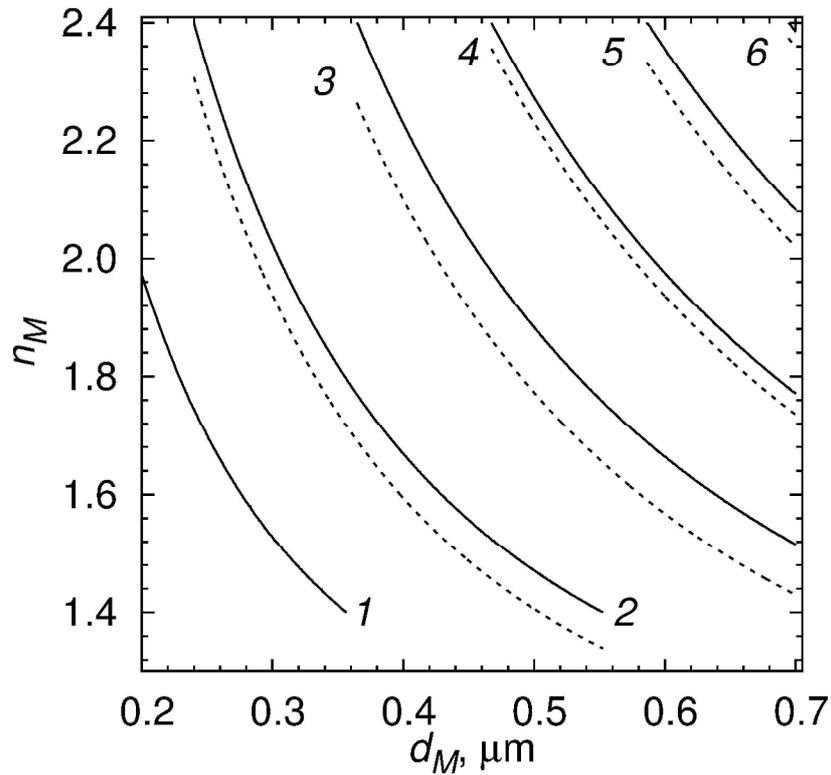

Figure 4. The solid lines match the valley bottoms (odd numbers) and the ridge crests (even numbers) as depicted in Fig. 2. The dotted lines correspond to the condition of the resonance (2 and 4) or antiresonance (3 and 5).

Comparing the positions of the loss maxima and minima with the resonance and antiresonance locations, we can conclude: the optical properties of the Bragg fiber with the intermediate layer of the sufficient optical thickness depend on the existence of the resonances and antiresonances in this layer.

## 5. CONCLUSIONS

In summary, we have shown the presence of the significant ranges of the intermediate layer parameters where the Bragg fiber with this layer is single-mode. With that, the guiding properties of the fiber substantially depend on the antiresonant reflection from the intermediate layer. Varying the refractive index of this layer, it is possible to significantly lower the optical loss. Near the resonances in the intermediate layer, there are several modes in the core of the fiber. The optical loss in this regime is reduced with an increase in the thickness of the intermediate layer. It is possible to control the optical properties of the fiber by changing only the thickness or refractive index of the intermediate layer.